\documentclass{article}
\usepackage{graphicx} 
\usepackage{authblk}
\usepackage{amsmath}
\usepackage{multirow}
\usepackage[sort, comma, numbers]{natbib}
\usepackage{pdflscape}
\usepackage[skip=5pt plus1pt, indent=40pt]{parskip}

\usepackage{geometry}
\usepackage{fancyhdr}
\title{Literary and Colloquial Tamil Dialect Identification}
\author[1]{M. Nanmalar}
\author[1]{P. Vijayalakshmi}
\author[2]{T. Nagarajan}
\affil[1]{Department of ECE, Sri Sivasubramaniya Nadar College of Engineering, Chennai, India.}
\affil[2]{Department of CSE, Shiv Nadar University Chennai, India.}
\date{}

\begin{document}
\maketitle

\section{Abstract}
Culture and language evolve together. With respect to Tamil, the form of
the language that people use nowadays has come a long way from its origins.
These days, the old literary form of Tamil is used
commonly for writing and the contemporary colloquial Tamil is used
for speaking. Human-computer interaction applications require 
Colloquial Tamil (CT) to make
it more accessible and easy for the everyday user and, 
it requires Literary Tamil (LT)
when information is needed in a formal written format. Continuing the
use of LT alongside CT in computer aided language learning applications
will both preserve LT, and provide ease of use via CT, at the same time.
Hence there is a need for the conversion between LT and CT dialects,
which demands as a first step, dialect identification. Dialect Identification
(DID) of LT and CT is an unexplored area of research. There is a
considerable research potential in this area, because, i) LT is
standardised while CT is not fully standardised and, ii) they have only
subtle differences. Five methods are explored in our work, which originated
from the preliminary work using Gaussian Mixture Model (GMM) 
for dialect identification of LT and
CT, which offered a motivation with an identification accuracy of 87\%. In
the current work, keeping the nuances of both these dialects in mind,
one other implicit method - Convolutional Neural Network (CNN); two
explicit methods - Parallel Phone Recognition (PPR) and Parallel Large
Vocabulary Continuous Speech Recognition (P-LVCSR); two versions of the
proposed explicit Unified Phone Recognition method (UPR-1 and UPR-2), are
explored. These methods vary based on: the need for annotated data, the
size of the unit, the way in which modelling is carried out, and the way
in which the final decision is made. Even though the average duration of
the test utterances is less - 4.9s for LT and 2.5s for CT - the systems
performed well, offering the following identification accuracies: 87.72\%
(GMM), 93.97\% (CNN), 89.24\% (PPR), 94.21\% (P-LVCSR), 88.57 (UPR-1),
93.53\% (UPR-1 with P-LVCSR), 94.55 (UPR-2), and 95.61\% (UPR-2 with
P-LVCSR).

\vspace{0.25cm}
\begin{center}
\textbf{Keywords}: \textit{dialect identification, literary and colloquial Tamil, implicit and explicit methods, statistical machine learning, neural network}
\end{center}

\section{Introduction }
\label{introduction}

Tamil is an ancient language that was chiseled to a state of beauty and
perfection. It is used by more than 80 million people around the world.
Apart from the state of Tamil Nadu, India, it is one
of the official languages of Sri Lanka and Singapore. There is a
significant Tamil speaking minority in Malaysia, Myanmar and the United
States of America.


With such a vast heritage and linguistic excellence, the language
evolved over time, but still retained its identity. The traditional
literary form of the language, from here on called Literary Tamil (LT),
possessed richness but is currently used only in the written form of the
language, in formal circumstances. The current colloquial form of the
language, from here on called Colloquial Tamil (CT), is widely used as a
way of vocal communication, but is not popular in the written form.
Hence, we are left with a need to preserve and develop the old and the
new respectively. In order to preserve and teach the richness of the
traditional literary form of the language, computers would need to
understand and process the literary form. In order to improve human
communication with computers, computers would need to understand the
colloquial form of the language. This naturally leads us to the task of
bridging the old and the new by finding ways to convert LT to CT and
vice-versa. To aid conversion, the input form of the language must first
be classified. When considering the differences and similarities between
LT and CT, they are quite analogous to two dialects. Hence, classifying
them can be considered a dialect identification task.

Although there is no specific definition of what colloquial Tamil should
be like (no standard grammar or lexicon), we define it as, `that which is
acceptable, understandable and neutral to the majority of Tamil
speakers'. It is the language that is used by the people from different
regions and social backgrounds and hence is popular in, and popularized
by, the media \citep{schiffman1999reference}.
This colloquialism is an
inevitable process that every language goes through and hence must be
addressed specifically.

Dialect Identification (DID) is an utterance-level variable-length
sequence classification task. We aim to extract the maximum information
out of a single utterance to make a decision. In one sense, DID is more
challenging than spoken Language Identification (LID) because of the lack of
much dissimilarity between dialects. 
It can also be considered a difficult case of language ID, 
where it is applied to a group of closely related languages 
that share a common character set \cite{zaidan2014arabic}.
Dialects are mostly similar, in that, 
there is a reasonable level of mutual intelligibility, but they also
vary in some characteristics. We aim to capture minute differences,
like attempting to distinguish between identical twins. 
Since the authors' native language is Tamil, they were able to 
learn the nuances of the different dialects of the language.

Existing work on Indian LID has not reached a critical level 
when compared to western and european languages \cite{aarti2018spoken}. 
While this is true for LID systems built on Indian languages, 
even though there exists a considerable number of them,
the fact is even more relevant for Indian dialects for 
which not much research interest exists at the moment.
This is due to the lack of datasets for Indian dialects.
For this reason, there aren’t many DID systems in Indian languages, 
especially Tamil, to which the current work 
could be directly compared to. 
But a comparison can be made with respect to dialect identification 
in other Indian languages, or LID systems in general.
Amongst the Dravidian languages --- Tamil, Telugu, Kannada 
and Malayalam --- DID systems for 
Telugu \cite{shivaprasad2020identification}, 
Kannada \cite{chittaragi2019automatic} and 
Malayalam \cite{sunija2016comparative}
have been addressed in literature.
The details of these systems are provided 
in Table \ref{tab:dravidian_dialects}.
The apparent need to address the dialects in Tamil is 
carried out in the current work considering a commonly 
spoken dialect in Tamil, namely colloquial Tamil.

\begin{table}[]
    \centering
    \caption{DID systems developed for Dravidian languages in literature.}
    \vspace{0.25cm}
    \begin{tabular}{p{3.0cm} p{2.5cm} p{2.5cm} p{2.5cm}}
        \hline
         & Telugu \cite{shivaprasad2020identification} & Kannada \cite{chittaragi2019automatic} & Malayalam \cite{sunija2016comparative} \\ 
        \hline
        No. of dialects & 3 & 5 & 2 \\ \hline
        Features & Mel Frequency Cepstral Coefficients & Spectral and Prosodic features & Spectral and Prosodic features \\ \hline 
        Classifier & Gaussian Mixture Model and Hidden Markov Model & Support Vector Machine and Neural Network & Artificial Neural Networks, Support Vector Machine and Naive Bayes \\ \hline
        Database & New database developed & New database developed & New database developed \\ \hline
        Best performance & 84.5\% & 99.24\% & 90.2\% \\ \hline
    \end{tabular}
    \label{tab:dravidian_dialects}
\end{table}

LID systems focussed on Indian languages, using various techniques, 
can be found in 
\cite{koolagudi2012identification, potla2018spoken, jothilakshmi2012hierarchical, reddy2013identification, shahina2005language}.
LID systems in which Tamil is one of the target languages can be found in
\cite{koolagudi2012identification, potla2018spoken, jothilakshmi2012hierarchical, reddy2013identification, matejka2005phonotactic, singer2003acoustic, lamel1994language, shahina2005language}.
When considering all these LID systems in literature, 
both generative (GMM or Hidden Markov Model (HMM)), and discriminative 
(Support Vector Machine (SVM) or Neural Network) models 
are found to be in use. 
The features used include, Mel Frequency Cepstral Coefficients (MFCC), 
prosodic features and Linear Predictive Coefficients (LPC). 
It should be pointed out that in almost all of these cases 
\cite{koolagudi2012identification, potla2018spoken, jothilakshmi2012hierarchical, reddy2013identification, lamel1994language, shahina2005language}, 
the LID performance of Tamil is highest amongst 
all the target languages included in the system. 
Similarly, from zissman’s comparison of four approaches \cite{zissman1996comparison}, 
we can see that Tamil is the only language which is not confused 
with other languages, when tested with both 45 and 10 second utterances.

From these evidences, it could be said of Tamil that it has 
distinct language cues which helps LID systems identify it accurately. 
But at the same time, in \cite{jothilakshmi2012hierarchical}, 
Tamil is confused with Malayalam and Telugu, 
which are both Dravidian languages. 
This seems to point to the fact that identifying Tamil is a 
tougher task when similar languages are included, and hence 
there is confusion. 
It would be fair to say that this confusion is even more so 
in the case of dialects. 
However, from the analysis and experiments in the current work, 
we find that the task of identifying literary and 
colloquial Tamil dialects is achievable with the help of 
conventional and proposed methods.

Further, in \cite{yanguas1999implications}, 
it was concluded that when the 
glottal flow derivative of two speakers were interchanged, 
the dialects were interchanged as well. 
This is because significant speaker and dialect specific information
such as noise, breathiness or aspiration, and 
vocalization is carried in the glottal signal. 
Based on these findings, it is suggested that 
even though a dialect identification task 
is usually considered to be closely associated
with the language identification task, 
it may actually be more closely related 
to the task of speaker identification.
Since MFCC is a feature that is commonly used
in speaker identification, it offers 
another justification for the use of MFCC 
in dialect identification as well. 
In \cite{pellegrino2000automatic}, 
it is suggested and proven that 
a significant part of the language is 
characterised by its vowels. 
This fact is in line with the proposal 
in the current work that, the nasal vowel carries 
one of the main cues that can distinguish between 
literary and colloquial dialects of Tamil.

The approaches to DID are similar to those used in LID \citep{etman2015language}.
Existing LID/DID systems can be classified into two major
categories, namely Explicit and Implicit LID/DID systems, based on whether
the system requires annotated corpora or not \citep{nagarajan2004language}.
Explicit methods require annotated corpus and hence are not usually
preferred if the aim is to identify many dialects or languages. However
in the current work, we address only two dialects and thus have the
advantage of working on both implicit and explicit systems. In this
context, conventional methods have been evaluated, with and without
modifications; and new methods have been proposed and evaluated.

Initially, implicit systems are built, because the speech data can be
used as such without annotation. In this regard, Gaussian Mixture Models
(GMM) and Convolutional Neural Networks (CNN)-based classifiers are built.
GMM-based classification \citep{etman2015language} is one 
of the simplest ways to identify dialects and the system offered a reasonable
performance. One of the reasons for good performance in this case 
may be the unique characteristic of CT - nasalized vowels \citep{nanmalar2019literary}.
Use of Gaussian mixture models for dialect/language identification 
is quite common in literature \cite{koolagudi2012identification, potla2018spoken, pellegrino2000automatic, jothilakshmi2012hierarchical, reddy2013identification, singer2003acoustic}. 
In \cite{koolagudi2012identification, pellegrino2000automatic, reddy2013identification}, 
GMM is the main modelling technique with MFCC, 
vowel information, and prosody being used as the main feature 
in these three systems respectively. 
In \cite{potla2018spoken}, 
GMM is compared with GMM-UBM (Universal Background Model), 
while in \cite{jothilakshmi2012hierarchical}, 
it is compared with HMM and Artificial Neural Networks (ANN). 
In \cite{singer2003acoustic}, 
it is combined with phone recognition and SVM. 
All these methods yield good performance in general, 
but it is agreed that even though GMM is 
successfully employed for speaker recognition, 
its language identification performance has 
consistently lagged that of phone-based approaches
\cite{singer2003acoustic}.

Generally, DID/LID does not perform well if
the data is noisy. The speech corpus used in the current work is
collected from scratch. 
Part of it is purposefully recorded in a mildly noisy environment,
to evaluate whether the systems work even in the presence
of noise. It is claimed that CNN works well on noisy data \citep{ganapathy2014robust}.
Hence a CNN-based system is developed and evaluated. 
Here, one dimensional(1D)-CNN is utilized.
A major motivation for using 1D-CNN is that 
the rhythmic and prosodic characteristic differences 
of literary and colloquial speech are reflected only across time. 
We believe that the network learns these patterns 
by convolving along time.
In \cite{baghel2019shouted}, 1D-CNN is used to learn features 
from the magnitude spectrum of speech frames in order to 
classify shouted speech from normal speech.
In the current work, we use MFCC features to train the network as in
\cite{moselhy2013lpc}. Whereas, \cite{ganapathy2014robust} and
\cite{najafian2018exploiting} extract features using CNN, but use other
methods such as GMM-ivector and SVM for classification respectively.
The manner in which 1D-CNN is adapted to learn from MFCC is detailed 
in Section \ref{dialect-identification-using-cnn}.

The simplest systems require only acoustic data, 
but in order to build better ones, 
phonetic transcription is also required \cite{pellegrino2000automatic}.
Working on only two dialects offered the motivation to develop some
explicit systems as well. In this regard, the annotated corpus, that is
required for explicit systems, is built. This task is particularly difficult
in the case of CT. The reasons for which will be discussed in Section
\ref{corpus}. The results are analysed and compared. The following
approaches and its variants are evaluated: Parallel Phone
Recognition (PPR), Parallel-Large Vocabulary Continuous Speech
Recognition (P-LVCSR), and Unified Phone Recognition (UPR).

Among the various approaches, phone-recognition approach offers
considerable promise, as it incorporates sufficient knowledge of the
phonology of the dialects to be identified \citep{ramasubramanian2003language}.
Considering the conventional explicit systems,
Phone Recognition Followed by Language Modeling (PRLM) and Parallel PRLM
methods are quite popular and are recommended when the transcribed data
is limited. On the other hand, if a good amount of transcribed data is
available for each dialect, Parallel Phone Recognition (PPR) can be
utilized \citep{etman2015language}. 
More use of PPR can be found in \cite{reddy2013identification} and \cite{lamel1994language}. 
In \cite{nagarajan2004language} and \cite{jayram2003language}, 
parallel recognition method, 
but with syllable-like and variable length subword units, 
instead of phones, are used. 
In \cite{lamel1994language}, 
a set of large phone-based ergodic HMMs 
are trained for each language and the language is 
identified as that associated with the model set 
having the highest acoustic likelihood. 
Here too, Tamil obtained the highest score. 
The issues in PPR for LID are studied 
in \cite{reddy2013identification}. 
In the current work, three different variations 
of PPR are built, and their performances evaluated.

So far, implicit systems with no memory (or temporal relationships) and
explicit systems with memory, but smaller units (phone) of knowledge,
were addressed. Application of higher knowledge sources to improve the
performance of DID follows. Full sentence recognition leads to better
performance than the same system using recognition of phonemes alone
\citep{hieronymus1996spoken}. 
A comparison of LID systems based on phone level 
and word level outputs with and without language model 
can be found in \cite{schultz1996lvcsr}. 
Here, it was demonstrated that the word-based system 
with trigram modelling of words offered 
superior performance than a phone-based system. 
Dragon systems has argued
that LVCSR is the best way to extract information from a given speech
signal \citep{mendoza1996automatic}. 
Even in the current work, this conclusion remained true and 
this method offered performance better than PPR 
and other implicit methods.

In addition to the methods above, we propose two methods based on LVCSR
named UPR-1 and UPR-2. These methods are
inspired from the P-LVCSR method, but the modelling technique and
decision making method is different. Unlike P-LVCSR, 
UPR-1 and UPR-2 use the
idea of universal modelling and word-based classification techniques. To
improve the performance of UPR-1, UPR-2 evaluates the word outputs again, to refine the
decision. These systems offer performance on par with the P-LVCSR
system.

The organisation of the paper is as follows. 
The nuances of literary and colloquial Tamil are detailed 
in Section \ref{nuances-of-literary-and-colloquial-tamil}.
The details of the corpus, both text and speech, are 
provided in Section \ref{corpus}. 
The six methods are split into three categories.
The first two methods, GMM and CNN, which are implicit methods,
are detailed in Section \ref{implicit-systems}.
The two explicit systems based on parallel recognition are 
detailed in Section \ref{explicit-systems}.
The two proposed systems based on unified phone modelling are
detailed in Section \ref{proposed-explicit-systems---upr}. 
The results are discussed in Section \ref{results-and-discussion}.

\section{Nuances of Literary and Colloquial Tamil}
\label{nuances-of-literary-and-colloquial-tamil}

A standard language can be thought of as possessing the following
functions, as per \cite{garvin1959standard}. They are,

\begin{itemize}
    \item The unifying function, which gives it the ability to unite multiple
      dialect communities into a single one that corresponds to the standard
      language.
    \item The separating function, which distinctly separates it from other
      standard languages.
    \item The prestige function, which gives the prestige of using the standard
      language.
    \item The frame of reference function, which gives it the ability to
      objectively define correctness of the language.
\end{itemize}

Literary Tamil used in the current work satisfies all the functions of a
standard language mentioned above. Over time, the literary form of Tamil
eventually became archaic. Due to the presence of an implicit need in
the society for a language that eases communication effort and is
acceptable by all, the standard language evolved to produce colloquial
Tamil. Some of the features of colloquial Tamil include the fact that it
does not contain features pertaining to caste, religion, or region. All
members of the community that can be identified with the standard form
of language, irrespective of the dialect (whether Madurai Tamil, Kongu
Tamil, Tirunelveli Tamil, Brahmin Tamil, etc.), understand and, most of
the time, speak the colloquial version of Tamil.

\begin{figure}[h!]
\centering
\includegraphics[width=0.90\textwidth]{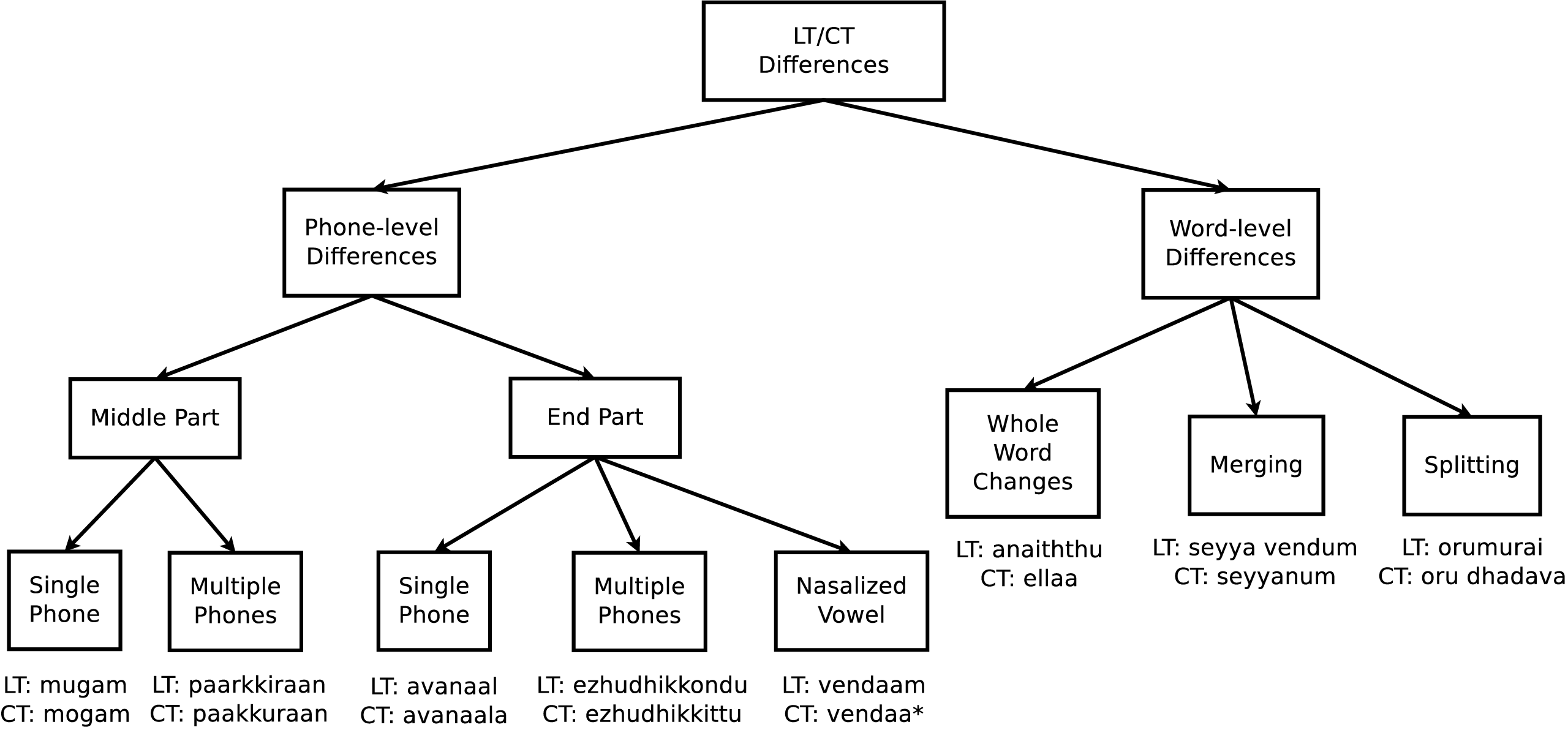}
\caption{Differences between literary and colloquial Tamil}
\label{fig:ltvsct}
\end{figure}

Due to urbanization and intellectualization, literary Tamil has taken
new forms, such as hyperlect (which is the Tamil used by the elite) or
edulect (which is the Tamil used by the educated), which can be thought
of as subclasses of colloquial Tamil. Irrespective of these recent
developments, the status of classical literary Tamil has not been
reached by the colloquial version. But, the problem lies in the fact
that, even though LT has a prestigious status, after a certain point of
time it was not used for oral communication. For informal oral
communication, CT is used.

Colloquial Tamil does not possess a standard orthographic transcription
or grammar. In \cite{james1985authority}, it is stated that ``In the strictest
sense, no spoken language can ever be fully standardized.'' He goes on
to say that writing and spelling are easily standardized, but the
standardization of the spoken form of a language is an ``ideology'' - an
idea, not a reality. With an absence of fully defined standards in
colloquial Tamil, \cite{joseph1987eloquence} talks about a restandardization
that can be expected in the future, since the language possesses
flexible stability. That is, a new version of the language with its own
spoken form, that challenges and attempts to capture some of the domains
of the older, can be expected. In \cite{schiffman1998standardization}, Shiffman
points out that colloquial Tamil, which he calls spoken Tamil,
has some set of standards. In the current work, we assume the same. The
assumption is that there is an implicit standard for colloquial Tamil which is
understood and accepted by most of Tamil speakers.

It must be noted that, there are a lot of similarities between literary
and colloquial Tamil, but the distinguishing factors, or the
differences, are comparatively less. Less because, the differences occur
mostly at the subword level where a phoneme or a small phoneme sequence
differs. Apart from this, there are also some cases when the whole word
is different. At the phoneme level, the change can be observed with one
or more phonemes. Usually this can be observed for the phonemes at the
middle and/or towards the end. At the word level, the whole word
differs. To clarify, the difference between the former and the latter
can be explained in terms of a total lack of correlation between the
phonemes of the equivalent literary and colloquial word. The differences
at the word level can also occur across a pair of words. Here, the two
words of one form (LT or CT) appear merged into one, in the other form;
or a single word splits into two.

One unique feature in colloquial Tamil is the acoustic property called
nasalization. Nasalization is the process in which a pair of phonemes in
a word containing a vowel followed by a nasal consonant, is replaced by the
nasalized version of the vowel alone. The end consonant is removed and
the vowel is called a nasalized vowel. In classifying literary and
colloquial Tamil, nasalization is a distinctive cue. Colloquial version
of Tamil trades `ease of speech' for `beauty and sophistication'. Two
examples follow. The phoneme (/zh/) in Tamil, adds beauty and
sophistication to literary Tamil. But it is replaced with its close
sounding relative phones: (/l/) or (/lx/) for the sake of ease
\citep{keane2006prominence}. Another example is that of the two phones
(/r/) and (/rx/) are considered seriously as possessing a difference,
but the difference is simply ignored in most colloquial speech. From the
above discussion we find that a lot of differences pertain to acoustic
differences in general, and phone level differences. It leads to the
conclusion that techniques that focus on these features work better.
More details about the similarities and differences between literary and
colloquial Tamil are provided in Figure \ref{fig:ltvsct}.

\section{Corpus}
\label{corpus}

The corpus used in the current work is created from scratch due to the
unique nature of the requirements. Both text and speech data in the
corpus are collected, recorded, and annotated, for the current task.

\begin{figure}[h!]
\centering
\includegraphics[width=0.97\textwidth]{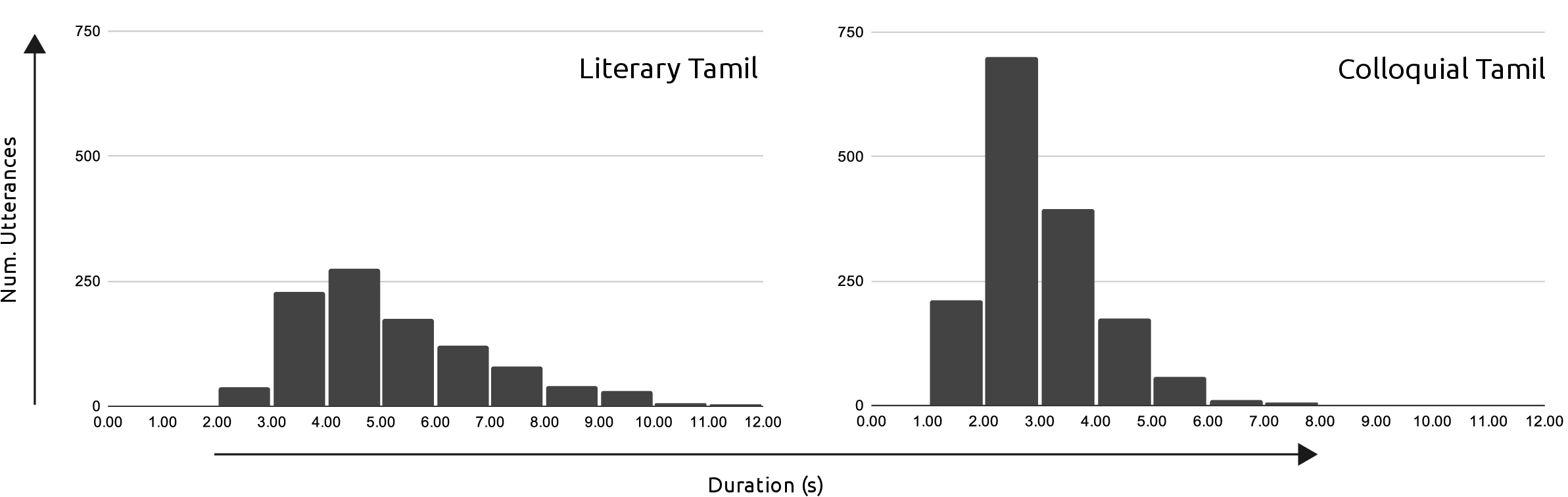}
\caption{Histogram of LT and CT Utterance Sizes}
\label{fig:LT_CT_Histogram}
\end{figure}

\subsection{Text Corpus }
\label{text-corpus}

The first source of both literary and colloquial Tamil text data was the
internet. Collecting literary Tamil text was simple. Since literary
Tamil has a standard grammar and orthographic structure, it is well used
in written form all over the internet in abundance. On the contrary,
colloquial Tamil's lenient standards make it difficult to have a
standard written form. For a long time, until a few decades ago, there
was no colloquial Tamil in literature. The earliest works using
colloquial Tamil include genres like novels and short stories. Many
writers focussed on reformation of the society took to such language to
depict the real life of people \citep{schiffman1998standardization}. But there
are some major problems. First is the lack of consistency of language
across writings from different authors. Each one followed their own
style and they also followed their own interpretation of colloquial
Tamil. Second, is code switching, by which, the authors drifted from one
dialect to another. Some writers may write purely in colloquial Tamil,
while others may mix it partly with literary Tamil, still others may use
region/religion-based dialects in between. Due to the factors mentioned
above, text data for colloquial Tamil is comparatively very much limited
and non-standardized on the internet. For the same reasons, colloquial
Tamil data collection and correction proved to be very difficult in the
current work. A lot of work was required in shaping the data into a
standard form. 
It must be noted that this is not a parallel corpus, since
manually collecting and correcting a parallel corpus is a tedious and
time consuming process which seemed to take away the focus of the
current work. It will be pursued in the future.

\subsection{Speech Corpus}
\label{speech-corpus}

Speech data is collected from volunteers who work inside the institute
of the authors. These volunteers are able to read and speak both
literary and colloquial Tamil text. Totally, 79 volunteers contributed
to this corpus. The speech data is collected from different age groups. These
include bachelor and graduate students, research scholars and faculties
of various ages. The audio is recorded using two different microphones
in parallel - a condenser and a dynamic microphone. The reason for this
is that the two have different frequency and gain characteristics. The
condenser microphone is sensitive and tends to pick up more from the
environment. The dynamic microphone is less sensitive to low gain and
high frequency and usually has a better signal to noise ratio when
compared to the condenser microphone, for the same environment. The
recording environment is a computer laboratory which is not sound
proofed. The noise levels are usually quite less, but with an air
conditioning unit or a ceiling fan switched on, the noise levels can be
higher. With an aim to record the speech in both noiseless and noisy
environments, to reflect real life data and for better generalization,
these appliances are switched on and off as needed for different sets of
data.

Some pre-processing steps are performed on these recordings. For the sake
of the comfort of the volunteers, the recording of each sentence in the
corpus is not time limited. This leads to the presence of leading and
trailing silences which need to be trimmed. Hence, silence trimming is
the first step. According to \cite{mendoza1996automatic}, trimming
silence is one of the simplest ways to obtain better results. The next
step is the preparation of the annotated corpus using phonetic
transcriptions and HMM-based viterbi alignment.
The corpus is then separated into train and test sets that comprise
approximately 70 and 30 percent of the total duration respectively.
During this separation, it is ensured that no speaker in the training
set is present in the testing set. Details of the corpus are provided in
Table \ref{table:corpus1} and Table \ref{table:corpus2}.
In all, 4387 LT and 5214 CT utterances were recorded.
Duration of the utterances in the
test set plays a big role in the results obtained 
\citep{etman2015language} \citep{muthusamy1994automatic}.
It measures the efficiency of the dialect
identification system. It is a measure of how well the system performs
with minimal evidence. Duration of all the test utterances in the corpus
is plotted as a histogram in Figure \ref{fig:LT_CT_Histogram}. The mean and variance of the duration of LT test utterances is 4.94(s) and 5.04(s) respectively and, for CT, it is 2.50(s) and 1.02(s) respectively.

\begin{table}[]
\centering
\caption{Details of the Speech Corpus - Male and Female speakers in Train and Test set}
\vspace{0.25cm}
\begin{tabular}{lcccc}
\hline
\multirow{2}{*}{} & \multicolumn{2}{c}{LT} & \multicolumn{2}{c}{CT} \\ \cline{2-5}
& Train  & Test  & Train  & Test  \\ \hline
\multicolumn{1}{c}{Total no. of Male Speakers}   & 21              & 3              & 20              & 6              \\
\multicolumn{1}{c}{Total no. of Female Speakers} & 43              & 12             & 37              & 15             \\
\hline
\end{tabular}
\label{table:corpus1}
\end{table}

\begin{table}[]
\centering
\caption{Details of the corpus. Duration of Train and Test utterances.}
\vspace{0.25cm}
\begin{tabular}{cccc}
\hline
& & LT & CT \\ 
\hline
\multirow{2}{*}{Train} & Num. Hours & 04:18 & 03:55 \\ 
 & Num. Utterances & 3371 & 3649 \\ 
\hline
\multirow{2}{*}{Test}  & Num. Hours & 01:23 & 01:51 \\ 
 & Num. Utterances & 1016 & 1565 \\
\hline
\end{tabular}
\label{table:corpus2}
\end{table}

\section{Implicit Systems}
\label{implicit-systems}
\subsection{Dialect Identification using GMM}
\label{dialect-identification-using-gmm}

A quick and effortless first level experiment to verify the validity of
the work was required since identification of colloquial Tamil is quite
an unexplored area of research. The experiment need not require
annotations and must validate the assumption that literary and
colloquial Tamil can be considered two dialects. This preliminary work
was required, because of the high similarities between LT and CT as
discussed in Section \ref{nuances-of-literary-and-colloquial-tamil}. 
The ability of dialect identification systems to capture these
fine differences, similar to a human being, was in question.
Implementing this preliminary system based on GMM with positive results
proved the proposed theory and enabled further work.

The Gaussian mixture models were built using 39 dimensional MFCC
features. The probability of a training feature vector $x_d$, belonging to a particular dialect (either literary or colloquial Tamil), being generated by a Gaussian mixture model that corresponds to the same dialect, can be given by, 

\begin{equation}
    p(x_d \mid \lambda_d) = \sum_{i=1}^M w_i g(x_d \mid \mu_i,\Sigma_i),
\end{equation}

\noindent where $\lambda_d$ is the model, which comprises the following for each mixture component, 
\begin{itemize}
    \item mixture weights $w_i$
    \item the means $\mu_i$
    \item the covariance $\Sigma_i$
\end{itemize}


The likelihood of a test utterance $X$ of duration $T$ is given by,
\begin{equation}
    L(X \mid \lambda_d) = \sum_{t=1}^{T} log\ p(x_t \mid \lambda_d),
\end{equation}
\noindent where $x_t$ denotes the feature frames of the utterance at time instants $t$ and, $d$ is the dialect which can take two values $lt$ and $ct$.

The maximum likelihood classifier's hypothesis of the dialect is given by,
\begin{equation}
    \hat{d} = arg\ \operatorname*{max}_{d = lt,ct}\ L(X \mid \lambda_d)
\end{equation} 

The number of mixtures were varied and the performance was
evaluated. A correctness measure of 87.72\% was obtained for 128 mixture
components. The conclusion of this experiment was that, even with many
similarities, the dialects have enough differences (like nasalized
vowels) to be classified with the help of just the acoustic features
\citep{nanmalar2019literary}.

\subsection{Dialect Identification using CNN}
\label{dialect-identification-using-cnn}

A one dimensional convolutional neural network (1D-CNN) is developed.
The convolutional neural network is not used to learn the
features, but rather learn the patterns in the input features that are
provided. The input feature is a 39 dimensional MFCC feature (13 Static
+ 13 Delta + 13 Acceleration). 
The number of timesteps/frames is to be kept
constant in a CNN, hence a fixed number is to be calculated. This is
carried out by computing the mean number of MFCC frames across all the
utterances in the corpus, and then adding a small constant value to it.
For the dataset in the current work, 
this is determined to be 440 frames, or 4.40s.
When the number of frames in a particular utterance is lesser than this
constant, zero padding is done, and when the size is more, the frames
are truncated. This way, a fixed number of timesteps for each
utterance is ensured. 
The constant is added 
to find the best trade-off between truncation and zero-padding.
For each sample, a categorical label that tells the
network whether the input sample belongs to LT/CT, is also provided for
training the network.

\begin{figure}[h!]
\centering
\includegraphics[width=\textwidth]{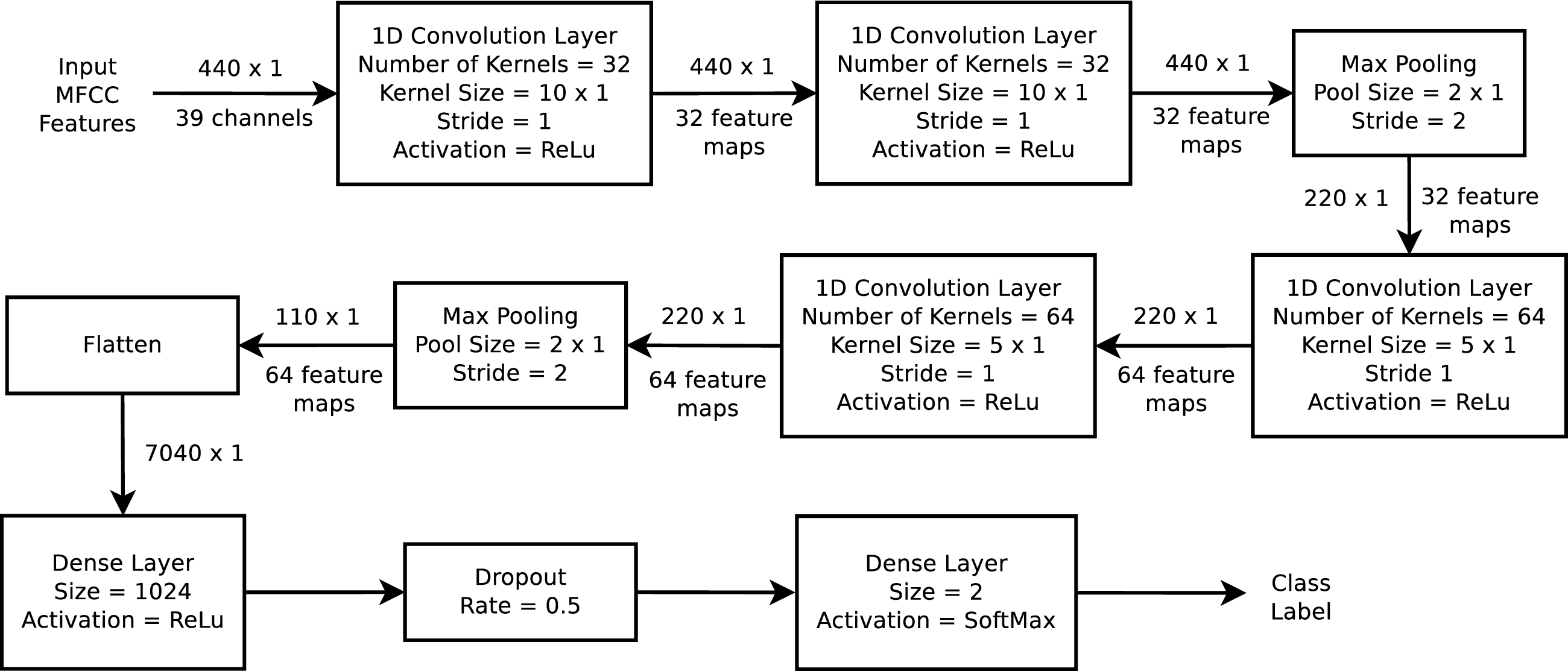}
\caption{Block Diagram of the proposed 1D-CNN architecture for LT/CT DID}
\label{fig:cnn}
\end{figure}

The use of 1D-CNN for this task is explained below.
Whether a CNN is 1D or 2D depends on 
whether the convolution operation is performed
across one or two dimensions.
In the current work, that dimension is 
only across time.
The 39 dimensions of the MFCC are assumed to be 
the channels of the signal. 
Hence, the dot product that produces each value 
in the feature map is two dimensional.
It can be given by,
\begin{equation}
    y_i^l = f(x^l \ast K_i^l + b_i^l)
\end{equation}
Where $x$ is the input, $K$ is the kernel, $b$ is the bias,
and y is the output of the $i^{th}$ kernel 
and $l^{th}$ layer. 
Here, $x$ and $K$ are actually two dimensional, where
one of the dimensions is the number of channels.
This is very similar to convolving a color image with
three channels, with a kernel that also contains a 
third dimension whose size is equal to 
the number of channels in the input.
For color images, the dot product is three dimensional.
Further details of the proposed architecture, 
can be found in block diagram shown in 
Figure \ref{fig:cnn}.
It shows the size of the input and the intermediate 
feature maps generated by the 1D-CNN.

The network is built using two sets of convolutional layers. Each
set contains two convolutional layers (totally four), 
a max-pooling layer and a dropout layer. 
The first two convolutional layers are made up of 32
filters with a kernel size of 10. The second set of convolutional layers
are made up of 64 filters with a kernel size of 5. A rectified linear
unit activation (ReLU) function is used in all the convolutional layers.
Zero padding is assumed in all convolutional layers to maintain 
the size of the input at each layer.
The pool size of the 1-dimensional max pooling layer is 2 and the rate
of the dropout is 0.25. The output of these convolutional layers are fed
to two fully connected dense layers, the last of which provides the
output. The first dense layer consists of 1024 outputs and is activated
by ReLU. The second (last) layer consists of the two outputs that we
require (LT/CT) and uses softmax activation. Using this setup an
identification accuracy of 93.97\% is obtained.

\section{Explicit Systems}
\label{explicit-systems}

Implicit systems mentioned so far learn from the speech directly without
having the need to possess sequential information or linguistic knowledge. 
On the other hand, explicit systems use linguistic knowledge
to build classifiers or identifiers. 
The phone recognizer present in all the methods that follow 
(PPR, P-LVCSR, UPR-1 and UPR-2) utilizes
39-dimensional MFCC features (13 static + 13 delta + 13 acceleration)
which worked quite well in the preliminary GMM experiments.
Cepstral mean subtraction was performed on the features to average out
the environmental differences in the recorded speech data.

\subsection{Dialect Identification using PPR}
\label{parallel-phone-recognition-ppr}
A parallel phone recognizer integrates the concept of phonology.
The PPR models all the phonemes
and builds a phone level language model for each dialect,
using bigram statistics obtained from the transcription in the dataset. 
This differentiates it from PRLM where phone recognizer output is used
to build the language model.

Two parallel recognizers are built - one for each dialect.
For each dialect $d$, there exists $N$ models 
$(\lambda_{d1}, \lambda_{d2}, ...., \lambda_{dN})$, 
that correspond to each of the phonemes. 
Here, each phone model is built with three states 
and the number of mixture components for the models 
is chosen based on the number of examples of each phone.
The set of models that correspond to all of the phonemes 
in a particular dialect is denoted as $\lambda_{dn}$. 
This differentiates it from the single model derived for the whole dialect, which is $\lambda_d$ mentioned in the previous section. It must be noted that the value of $N$ varies for each dialect, based on the number of phonemes present in the dialect.

\begin{figure}[h!]
\centering
\includegraphics[width=0.70\textwidth]{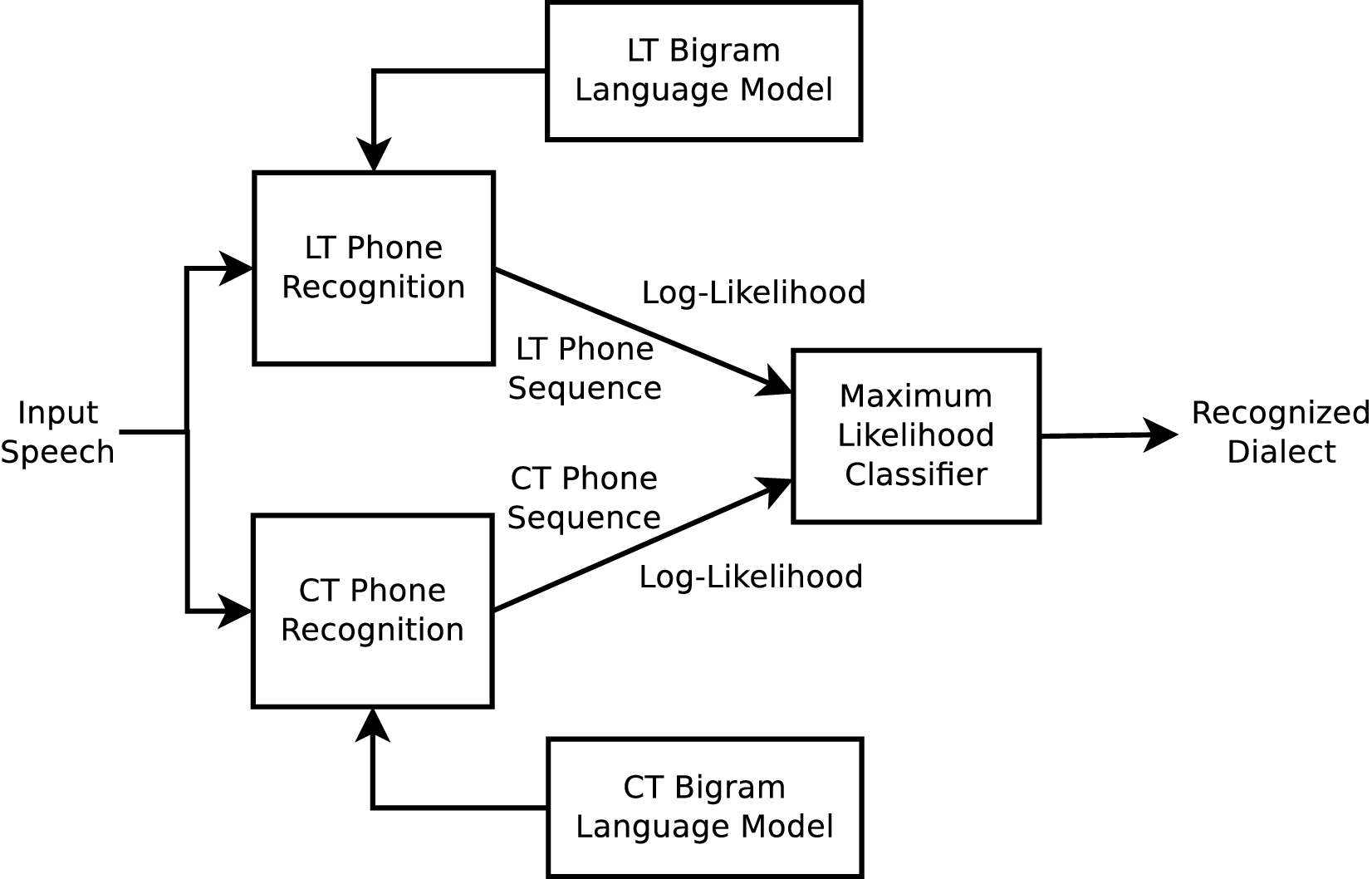}
\caption{Block Diagram of a PPR System}
\label{fig:ppr}
\end{figure}

For each test utterance, both recognizers are used to produce a phone
sequence and its corresponding likelihood. The dialect $\hat{d}$ is predicted using Viterbi decoding to decode the sequence of phonemes, by computing the log-likelihood score corresponding to the best path through the recognizer that belongs to each dialect,

\begin{equation}
    \hat{d} = arg\ \operatorname*{max}_{d = lt,ct}\ 
    L(\hat{p}_d \mid \lambda_{dn}),
\end{equation}

\noindent where $L(\hat{p}_d \mid \lambda_{dn})$ denotes the log likelihood of the viterbi path $\hat{p}_d$ for dialect $d$. The accumulated likelihood scores are
then used to identify the dialect. This process is illustrated in the
block diagram shown in Figure \ref{fig:ppr}. PPR is good at language
identification since, when it comes to languages, there are 
considerable differences in the acoustic realization of the same phones.
In addition to this,
they also possess unique phones that differentiate it even further.
However in the current scenario, the two dialects in focus have many
(almost all) common phones. To explain further, LT has a total of 39
phones and CT has 40 phones. The 39 phones in LT are common to both LT
and CT. The one unique phone in CT is a by-product of code switching of
CT to English. With this kind of a setup, the authors were skeptical
about the application of PPR for the current work. But it still worked
and yielded good performance. The reasons maybe the following,

\begin{itemize}
\item
  Word endings: Colloquial Tamil words mostly end in a vowel, when
  compared to LT, where only a few words contain this feature. It
  can be hypothesized that the vowels in these two dialects are modelled
  differently because of this factor. Vowels being an important
  distinguishing factor, could play a major role in classification.

\item
  Effort in Speaking: There is a difference in the effort used by people
  to speak LT and CT. Native Tamil speakers speak CT effortlessly,
  while LT is spoken more rigidly.
  This creates the following differences:

  \begin{itemize}
  \item
    Duration: There is a slight difference in the duration of similar
    sentences spoken in LT and CT. CT sentences tend to be slightly
    shorter.
    
  \item
    Vowel Duration: The duration of the vowels specifically is reduced
    in CT.
    
  \item
    Stress: LT being a formal language, requires the speakers to stress
    certain phones the way it is meant to be. CT maintains its casual
    approach and aims at efficiency than beauty. This aspect is explained
    with examples in Section \ref{nuances-of-literary-and-colloquial-tamil}.
    
  \item
    Rhythm: One of the beauties of LT is its tendency to be rhythmic. CT
    does not have this feature and tends to be more spontaneous and free
    flowing.
    
  \end{itemize}
\end{itemize}

These reasons help us understand why PPR worked, and why phone-based
ideas in general will work.

\subsubsection{Version 1 - Conventional PPR}
\label{version-1---conventional-ppr}

A conventional PPR system, that contains both phone recognition and
language modelling, was built. A language
model that contains both unigram and bigram statistics of the phones is
used. The identification efficiency/performance of this method is
85.24\%. Considering the best scores of each method 
mentioned in the current work, GMM obtained the least score.
Even though the score obtained with the first version of this system is the
least amongst the PPR-based systems and is also lower than the 
score obtained using GMM; performance better than GMM
can be obtained using slight modifications to this system,
as can be seen in the sections that follow.

\subsubsection{Version 2 - Unique Phone Inclusion}
\label{version-2---unique-phone-inclusion}

The reason for the slightly lesser performance of PPR - Version 1 is the
lack of unique phones. Almost all the phones between LT and CT are
similar. This problem can be addressed by highlighting/specifying some
unique characteristics in the phones. As discussed in Section \ref{nuances-of-literary-and-colloquial-tamil}, CT has a unique
characteristic in vowel sounds, called nasalization. Hypothetically, if
the effect of nasalization can be understood and incorporated in the
system, results could be boosted. In \cite{alorifi2008automatic}, it is
concluded that specifying dialect unique sounds will allow the DID to
classify both dialects accurately. Hence, in this version, the
occurrences of nasalized vowels were located. 
This was carried out using the characteristics of nasalized vowels
detailed in Section \ref{nuances-of-literary-and-colloquial-tamil}.
Irrespective of the individual vowel, all nasalized vowels which come
under this category were identified, grouped together and modelled. This
unique model from CT provided an increase in performance as expected,
with an identification score of 88.60\%.

\subsubsection{Version 3 - Language Model Exclusion}
\label{version-3---language-model-exclusion}

The conventional PPR system mentioned in 
Section \ref{version-1---conventional-ppr}, 
integrates the scores from acoustic modelling and language
modelling to compute the final score. This is called
acoustic-phonotactic decoding or simply, joint decoding \citep{zissman1996comparison}.
This is done because the inclusion of language/dialect
specific knowledge will make the system a better identifier. But here, in
our case, the problem is the presence of similar phonemes and partly
similar syntax between the two dialects. This factor introduces overlap
in the corresponding language models which lead to lower performance. In
this version, the decision is made using the scores from the acoustic
models alone. As mentioned in Section 
\ref{nuances-of-literary-and-colloquial-tamil}, even though we
have similar phone sets for both dialects, there are a lot of unique
parameters which work implicitly and contribute to the DID process. This
approach provided a further raise in performance, with an identification
score of 89.24\%, because acoustic data is the major source of information
in speech.

\subsection{Dialect Identification using P-LVCSR}
\label{p-lvcsr}

Phonemes are smaller units and most of the differences at
the phoneme level were exploited so far. To increase the performance
further, higher level units like words and word sequences can be used to
improve the identification results. These higher level units provide
more evidence. 
In \cite{arslan1996language}, the authors claim that when we deal
with local dialects, working with isolated words is better than working
with phone models. However in the current work, we identify the dialect
from continuous speech. In this experiment, along with the advantages of
phoneme modelling, the advantages of using a word-based lexicon is also
utilized. P-LVCSR is similar to PPR, but the likelihood is conditioned by both the phone model and the list of possible words ($W$) in the dialect.

\begin{equation}
    \hat{d} = arg\ \operatorname*{max}_{d = lt,ct}\ \sum_{t=1}^{T} L(\hat{p}_d \mid \lambda_{dn},W)
\end{equation}

\begin{figure}[h!]
\centering
\includegraphics[width=0.72\textwidth]{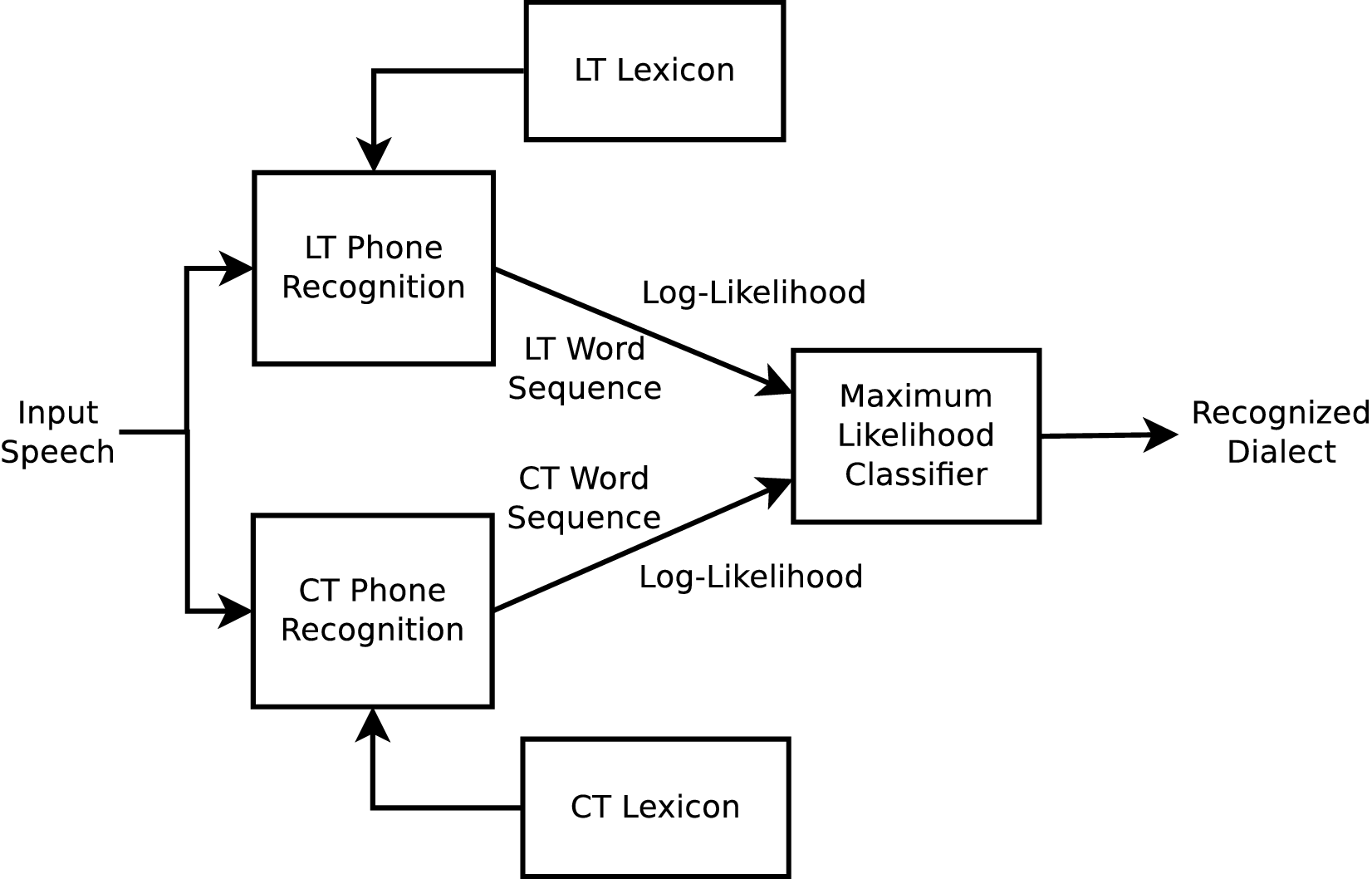}
\caption{Block Diagram of a P-LVCSR System}
\label{fig:plvcsr}
\end{figure}

As mentioned in \cite{etman2015language}, word could be chosen as the unit
of DID, which in turn can be divided into strings, in our case phonemes.
In this scenario, the phone models capture all the fine details, while
the word lexicon helps the system to obtain the larger picture by
providing the sequence of phones in that word. If some of the phones in
a particular word is recognized wrongly, the word-based lexicon helps
the system to identify the right word despite the errors. The block
diagram in Figure \ref{fig:plvcsr} represents the architecture of this method.

In Figure \ref{fig:plvcsr}, the phoneme recognizer is built using triphone models. 
During model creation, similar triphones were tied together and then mixture
splitting was performed for 4, 8, 12 and 16 mixtures. Best results were
obtained for 16 mixtures. 
This model configuration is maintained 
in the subsequent methods that follow (UPR-1 and UPR-2).
The total number of triphones created from
context independent phones are 3488 for LT and 3995 for CT. Lexicons are
created for both dialects. LT lexicon contains 4879 entries whereas CT
contains 4355.

From Figure \ref{fig:plvcsr}, we can see that if an unknown test utterance is
given, it is processed by both phone recognizers in parallel and, using
the lexicon, a word sequence along with its accumulated likelihood is
generated by each of the recognizers. The sequence which has the maximum
likelihood is chosen as the output sequence. A point to be mentioned
here from \cite{lowe1994language} is the claim that this method is a success
because it works even in domains in which the recognition task, of
acquiring accurate transcriptions, is too difficult. Similarly, the
recognizers in the current work do not have commendable performance.
They provide meaningless word sequences even if it is from the
recognizer belonging to the right dialect. Still, P-LVCSR is a
successful method for the current task because it focuses on finding the
difference between the dialects rather than focusing on recognizing the
words \cite{hieronymus1996spoken} \cite{mendoza1996automatic} \cite{lowe1994language} \cite{schultz1996lvcsr}.
It must be mentioned here that language models
were not included in this method. The reason being that the amount of
useful data for CT is limited, which the authors aim to resolve in the
future. The identification accuracy of this method is 94.21\%, which is
higher than PPR, GMM and CNN.

\section{Proposed Explicit Systems - UPR}
\label{proposed-explicit-systems---upr}

\subsection{Motivation}
\label{proposed-explicit-systems---upr-motivation}

All the methods mentioned above were originally developed to address the
LID problem, but they can be adopted to the DID problem as well due to
many similarities between both. Our experiments also prove that these
methods work well for the case of literary and colloquial Tamil. But
since we have knowledge of the dialects in focus, rather than adopting
only generalized solutions, more fine tuned approaches can also be
utilized. In PPR and P-LVCSR methods, models and lexicons are built
separately for LT and CT. If a test utterance is provided to the system,
the LT models and/or lexicon and the CT models and/or lexicon compete to
provide two different phone/word sequences as outputs, with their
corresponding likelihoods. Based on these outputs the corresponding
dialect is identified.

The UPR method arises out of the intuition of having a unified system
with a unified set of models and unified output.
While the previous methods focused on the differences between
LT and CT, the proposed method focuses on the commonness.
To provide an analogy:
the acoustic characteristics of literary and colloquial Tamil can be
likened to two versions of the same car released at different time
periods. Although they would differ in a few ways, namely in new
features being added to, and some old features being removed from, the
new version of the car; it would still essentially be the same car.
Focusing such similarity/commonness between literary (old) and
colloquial (new) Tamil into account, a unified phone recognition makes
sense. For example, considering the phoneme /a/: both literary and
colloquial Tamil have their own acoustic characteristic since it plays a
different role in the corresponding dialects. 
Despite these differences, we can find considerable commonality 
between them and model them together. 
In the same way, the lexicon is also combined. 
That is, LT and CT words are put together to create
a single unified dictionary.

A major advantage of using such a unified system is that, 
when the data consists
of utterances that have both LT and CT words together, instead of having
exclusively LT or CT words, they can be modelled as such and
considerable performance can be achieved even in this scenario.
Although it is simple to collect LT text data from the web, 
as mentioned in Section \ref{text-corpus}, it is not always 
possible to collect `pure' LT text data. This is due to the influence of 
colloquial Tamil in these modern day sources which are deemed to be
written in literary Tamil.
Similarly, and sometimes more severely, CT text data also has 
the same problem, as mentioned in Section \ref{text-corpus}.
Public data from the internet (facebook, twitter, youtube comments, etc.) 
and some literature contain mixed data. 
Obtaining pure LT or CT utterances from these sources through 
classification and conversion using 
a natural language processing-based algorithm is not a trivial task. 
Besides, the algorithm itself will have its own shortcomings and
this would be reflected in the processed data.
The same is conveyed in \cite{zaidan2014arabic} where, 
the authors working on arabic text dialect identification suggest that, 
even though dialects can be considered as two languages, 
it is not a trivial matter to automatically distinguish them.
In the current work, to prevent such disorderliness in the data,
for use in the explicit methods, considerable efforts were taken 
to pre-process the text manually, before recording the utterances.
This is to make sure that each sentence is either pure LT or CT text.
When working with a large corpus, 
proportionally more work would be required.
However, the proposed method is a good solution for 
the issue discussed above. 
Here, pre-processing mixed text is not required, 
since the core intuition of the method is unification. 

Here, all the 39 phones that are common to both LT and CT are modelled
together, while the one unique phone in CT is modelled separately.
Together, the unified phone recognition is performed with 40 models.
Since this system contains only one recognizer, both the LT and CT
lexicons are combined, just like how the phones were modelled together
and produces only one output sequence. The dialect is identified from
this sequence. The log likelihood of the sequence of words $W$,
given the unified phone model $\lambda_u$ and unified word list $W_u$ is given as,
\begin{equation}
    L(W \mid \lambda_u, W_u) = \sum_{t=1}^{T} log\ p(x_t \mid \lambda_u,W_u).
\end{equation}

\begin{table}[]
\centering
\caption{Simplified example of UPR output sequences demonstrating three cases of bias. $\text{W}_i$ represents words in the output sequence.}
\vspace{0.25cm}
\begin{tabular}{lllllll}
\hline
 & $\text{W}_1$ & $\text{W}_2$ & $\text{W}_3$ & $\text{W}_4$ & $\text{W}_5$ & Bias \\ \hline
Case 1               & LT & LT & LT & CT & LT & Biased towards LT \\
Case 2               & CT & LT & CT & CT & LT & Biased towards CT \\
Case 3               & LT & CT & CT & LT & & Equi-probable \\ 
\hline
\end{tabular}
\label{table:upr_output}
\end{table}

\begin{figure}[h!]
\centering
\includegraphics[width=0.95\textwidth]{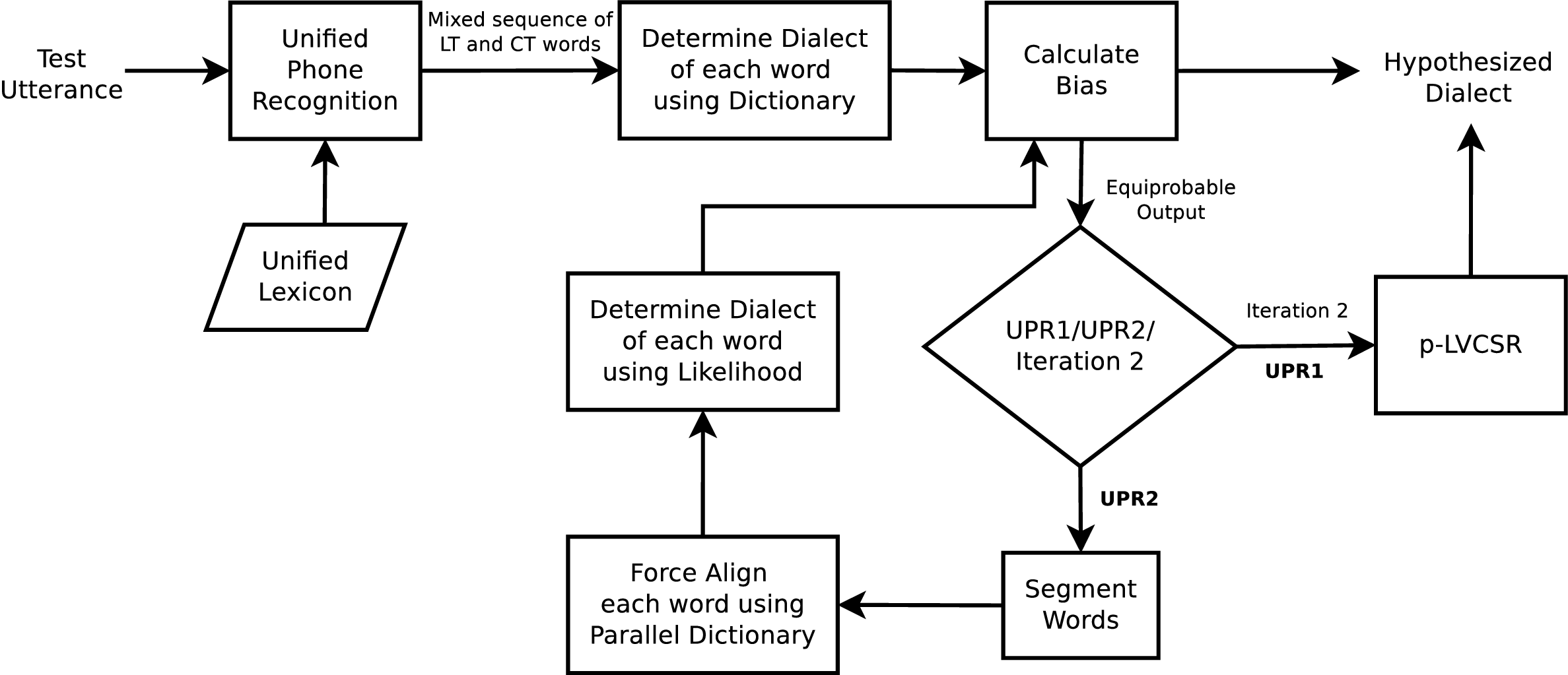}
\caption{Block Diagram of the proposed UPR system showing both variations}
\label{fig:upr}
\end{figure}

Unique words between both LT and CT are retained in the
unified lexicon. The unified lexicon contains 8907 entries in total. For
recognition, both phone models and the lexicon play an equally important
role. The output produced by the system is usually a meaningless sequence of
literary and colloquial Tamil words. 
The dialect identification is
performed based on the bias, which is given by, 
\begin{equation}
    \hat{d} = arg\ \operatorname*{max}_{d = lt,ct} count(W_d)
\end{equation}
where, $W_d$ is the number of words that belong to each dialect, 
and the dialect is determined using separate dictionaries/lexicons.
A simplified example of this UPR output sequence is provided 
in Table \ref{table:upr_output}.
For an output sequence, if the
number of words that belong to CT is more than that of LT, then it is
biased towards CT, and hence the sentence will be identified as CT. If
the sentence contains an equal number of LT and CT words, then it is
considered as an equiprobable case and, is handled separately as a
special case. The reason for this scenario could be i) inefficiency or,
low performance of the recognition module ii) confusions due to combined
lexicon. How this special case is handled, determines whether the system
belongs to UPR-1 or UPR-2. The block diagram in Figure \ref{fig:upr} explains
these methods in detail.

\subsection{Dialect Identification using UPR-1}
\label{unified-phone-recognition---type-1}

In UPR-1, the special equiprobable case, mentioned in 
Section \ref{proposed-explicit-systems---upr-motivation},
is handled by utilizing the
classification outputs of the P-LVCSR. 
In \cite{lloyd1998recurrent}, it was concluded that fusing the results
of their own method with that of an existing method provided better
results than using just their method in isolation. In the same way, the
identification accuracy of UPR-1 in isolation is 88.57\%, but together
with P-LVCSR the accuracy increases to 93.53\%.

\subsection{Dialect Identification using UPR-2}
\label{unified-phone-recognition---type-2}

In UPR-1, a work around is used to solve the equiprobable constraint by
using the results from P-LVCSR. But in UPR-2, the issue is addressed
directly. The intuition behind this approach is that another round of
recognition, i) from a different perspective and ii) with new
knowledge, would refine the output. In this way, hopefully, a bias is
added, the equilibrium is removed, and the dialect can be identified.

For example, assume a hypothetical equiprobable case where the result
hinges on correctly identifying one of the words (say, target word),
that was originally incorrectly identified by the unified phone
recognizer. A better decision can be reached if each of the words,
including the target word, undergoes a reconfirmation process. 
Here, the words are segmented and processed using the
information from the word label outputs obtained from the unified phone
recognizer. The steps of the reconfirmation process is as follows:

\begin{enumerate}
\def\labelenumi{\arabic{enumi}.}
\item
  
  Find the corresponding parallel word.
  
\item
  
  Phonetically align both the recognized and its parallel word using 
  HMM-based forced Viterbi alignment.
  
\item
  
  The dialect of the word that has the maximum likelihood is chosen.
  
\end{enumerate}

For each word segment $x_{wt}$ in the utterance, and its estimated word $w$,
find the likelihood of the sequence of phones in that word using Viterbi alignment,
with respect to the same word and its parallel word in the other dialect.
Here $w_d$ is the word from each dialect.

\begin{equation}
    L(w \mid x_{wt},w_{d}) = \sum_{t=1}^{T} log\ p(x_{wt} \mid \lambda_u, w_{d})
\end{equation}

Dialect of the word is given by, 
\begin{equation}
    \hat{d} = arg\ \operatorname*{max}_{d = lt,ct} L(w \mid x_{wt},w_{d})
\end{equation}

Doing this, would change the dialect of the target word hence shifting
the bias towards the actual dialect of the utterance.

In order to obtain the parallel word, a parallel dictionary is required.
One that finds the corresponding CT word for an LT word and vice versa.
This dictionary was created using linguistic conversion rules and
manual conversion, due to the absence of a parallel corpus. 
Based on the phonotactic rules available in 
\cite{schiffman1999reference}, 
LT to CT conversion was possible to a certain degree. 
But the reverse is not a trivial task since there 
is no established standard, hence CT to LT conversion 
was carried out manually.
Like in all cases, here too, manual conversion is a
time consuming process. 
This is a drawback of this method. 
Yet, UPR-2 can be made fully automatic if parallel corpus is available. 
To create one such corpus, considerable research effort is required. 
The authors are interested in pursuing this in the future. 
The scope of the current work is to showcase a 
proof of concept for the proposed method.

In some cases, the output still remains
equiprobable. Although this is a very rare scenario, it is solved by
using P-LVCSR output as in UPR-1. A block diagram of this process is
shown in Figure \ref{fig:upr}. There is close to a 4\% performance
difference between the vanilla UPR-1 (88.57\%) and UPR-2 (92.29\%) methods,
without the P-LVCSR step. It must be noted that, in this scenario, an
equiprobable case is considered a negative result.

The following experiments were carried out on the basic unified phone
recognition system - without P-LVCSR, to resolve issues that could
further improve its performance.

\subsubsection{Issue 1 - Same Parallel Word}
\label{issue-1---same-parallel-word}

In this case, the parallel word derived from the dictionary is the same
as the recognized word. There are two ways to handle this. Solution 1:
The class of the recognized word is retained as the final class of the
word, which is the default method. Solution 2: The class of the
recognized word is ignored and is not considered for the computation of
the bias. The word `enna' which is common in both LT and CT falls under
this scenario. Using \emph{Solution 2}, an improvement of about 2\% was
obtained. From 92.29\% for basic UPR-2 to 94.07\%.

\subsubsection{Issue 2 - Alignment Unsuccessful}
\label{issue-2---alignment-unsuccessful}

In some cases the derived parallel word, corresponding to the recognized
word, is much longer than the recognized word. Based on the length of
the original utterance, it may not be possible to align the derived
parallel word with it, and no likelihood value would be obtained. Hence,
in this case, the dialect of the recognized word is retained. Example:
`paattutu' (CT, recognized) and `paartukkondeu' (LT, derived from
parallel dictionary). Here the latter is much longer than the former and
hence can cause errors during alignment. Using this solution improved
the result from 94.07\% to 94.40\%.

\subsubsection{Issue 3 - Inappropriate Parallel Word}
\label{issue-3---inappropriate-parallel-word}

In some cases, the parallel word in the dictionary may not be
appropriate because of rule-based conversions. In such cases, the only
option is to manually identify these problematic words and make the
necessary corrections. Example: `wiyanddanar' (LT, recognized) and
`weyanddanar' (CT, derived from parallel dictionary). Here, the
appropriate CT word is `wiyandhaanga'. Making such manual changes
improved the result from 94.40\% to 94.55\%.

After identifying these issues and fixing them, utilizing P-LVCSR
offered a final improved result of 95.61\% which is a near 1\%
improvement. This shows that both UPR-1 and UPR-2 are considerably
successful in identifying two similar dialects: LT and CT. If more
efforts are made towards improving the performance of the LVCSR and
building parallel dictionaries, much better identification accuracies
can be achieved.

\section{Results and Discussion}
\label{results-and-discussion}

\begin{table}[]
\centering
\caption{Identification Accuracies of each method}
\vspace{0.25cm}
\begin{tabular}{cc}
\hline
    & Identification Accuracy \\ \hline
GMM     & 87.72\%                          \\ 
CNN     & 93.97\%                          \\ 
PPR     & 89.24\%                          \\ 
P-LVCSR & 94.21\%                          \\ 
UPR-1    & 93.53\%                          \\ 
UPR-2    & 95.61\%                          \\ \hline
\end{tabular}
\label{table:results}
\end{table}

The results of all the methods discussed are consolidated in Table
\ref{table:results}. GMM being a preliminary and easy to develop implicit system,
was the first method experimented. It offered a good score, but in
comparison it is the lowest amongst all methods. Similar to GMM, CNN,
although being a recently developed technique, does not require
phonetically segmented data. It uses only acoustic features for
learning, but the performance is better in comparison. All the four
methods that follow require phonetic transcriptions since they model the
individual phones. PPR offered the lowest performance amongst the
explicit methods. Reason being the subtle differences between the phones
of the two dialects. P-LVCSR, which is quite similar to PPR, offered
better performance. This is due to i) word outputs in P-LVCSR providing
better difference in accumulated likelihoods compared to phone outputs
in PPR and, ii) higher level units as outputs in P-LVCSR uses more
evidence from the language. 
Among the
two variants of the proposed UPR method, which used a unified modelling
scheme for both the dialects, UPR-1 on its own, without the P-LVCSR
refinement, offered a performance that was close to GMM and PPR. The
basic score of UPR-2 (without P-LVCSR), is much higher than that of UPR-1.
It can also be seen that while P-LVCSR offers considerable improvement
in performance in the case of UPR-1, it offers negligible performance
improvement in the case of UPR-2, which speaks for the basic technique of
UPR-2. Most of the methods discussed in the current work
offer performances that are quite close to each other. 
The effects of these differences may not be visible
in a real world scenario.

\begin{table}[]
\centering
\caption{Confusions in each method}

\vspace{0.25cm}
\begin{tabular}{ccccc}
\hline
\multicolumn{1}{l}{} & LT as LT & LT as CT & CT as LT & CT as CT \\ \hline
GMM           & 0.88              & 0.12              & 0.13              & 0.87              \\
CNN           & 0.90              & 0.10              & 0.02              & 0.98              \\
PPR           & 0.89              & 0.11              & 0.11              & 0.89              \\
P-LVCSR       & 0.95              & 0.05              & 0.06              & 0.94              \\
UPR-1          & 0.96              & 0.04              & 0.09              & 0.91              \\
UPR-2          & 0.97              & 0.03              & 0.06              & 0.94              \\ \hline
\end{tabular}



\label{table:confusion_matrix}
\end{table}

Table \ref{table:confusion_matrix} shows the confusions in all six techniques. 
From the table we can infer that, 
a lot of confusions happen for colloquial Tamil. The reason
is that, the average duration of a CT utterance is shorter in comparison
\citep{niesler2006language}. If utterances with similar lengths, or a
parallel corpus is available, this factor could be overlooked and a
better analysis could be provided. An unique result can be observed for
the case of CNN. Here we see that CNN handles CT better than all other
methods. This could be because of the fact that CNNs learn the full
utterance as a whole, rather than learning it frame by frame or, one
phone label after the other as in the case of GMM and the other methods
respectively. This process could have captured the wholesome acoustic
characteristics of the dialect. Another probable reason is that while
pre-processing the utterances for CNN, the size of the data is kept
constant by zero padding or truncating it. Hence removing the bias that
the length of the sentence offers, as mentioned earlier in the case of
other methods.

\section{Conclusion}
\label{conclusion}

Dialect identification is impactful to both academia and the industry,
as it can create a positive impact in the society. Colloquial Tamil and
the other dialects of Tamil comprise an interesting area suitable for
research exploration. In the current work, the DID of two similar
dialects - with only subtle differences - in Tamil (LT and CT), using
six different methods, are explored. These include two implicit (GMM and
CNN), two explicit (PPR and P-LVCSR) and two proposed explicit systems
(UPR-1 and UPR-2). A database of considerable size is collected from
participants of different age groups and genders, using different
microphones and in different environments. GMM and CNN being implicit
methods, were easy to develop and they offer considerable performance.
CNN works uniquely well for CT. PPR and P-LVCSR, being
explicit systems, required annotation of the speech data. Two
recognizers, for LT and CT, were developed for each technique. The
performance of PPR is not comparable to the effort it takes to annotate
the data, while P-LVCSR offered considerable performance,
being a word-based system. The proposed methods, UPR-1 and UPR-2 which
focus on the commonality between the dialects, are advantageous when
organized data is not available. That is, if the recorded utterances
contain multiple dialects (with similar phone-sets) together within the
same utterance, UPR methods can be used without compromise. While UPR-1
is a basic method relying on the bias in the utterance, UPR-2 utilizes
another level of confirmation using HMM-based forced Viterbi alignment,
to identify the dialect. While the popular opinion in literature is that
word-based methods offer best results, phone-based methods and methods
based on only acoustic information also work well for the current task.
When there is a
limitation in the availability of a complete lexicon, phone-based
systems can be used. When organized data is not available, then UPR
methods can be used. If using language information is not desired,
to eliminate manual efforts and computational costs,
implicit methods can be used. Despite differences in the approach, all
methods offered decent performances that are quite close to each other
(around 90\%).

\end{document}